\documentclass[letterpaper, journal]{IEEEtran}
\IEEEoverridecommandlockouts
\usepackage{cite}
\usepackage{amsmath,amssymb,amsfonts}
\usepackage{algorithmic}
\usepackage{graphicx}
\usepackage{textcomp}
\usepackage[dvipsnames,svgnames]{xcolor}
\usepackage{pbalance}
\usepackage{siunitx}
\usepackage[switch]{lineno}
\usepackage[sort=none]{glossaries}
\usepackage{multirow}
\usepackage{booktabs}
\usepackage{tabularx}
\usepackage{todonotes}
\usepackage{csquotes}
\usepackage{multirow}
\usepackage{caption}
\usepackage[position=b]{subcaption}

\usepackage[hidelinks]{hyperref}

\newif\ifanonymous

\let\orgautoref\autoref
\renewcommand{\autoref}
{\def\sectionautorefname{Section}\def\subsectionautorefname{Section}\def\subsubsectionautorefname{Section}\def\figureautorefname{Fig.}\def\equationautorefname{Eq.}\orgautoref}

\usepackage{xspace}
\newcommand{\etal}{\textit{et al.}~}

\newcommand{\ie}{\textit{i.e.,}~}
\newcommand{\cf}{\textit{cf.}~}

\newcommand{\one}{({\em i})\xspace}
\newcommand{\two}{({\em ii})\xspace}
\newcommand{\three}{({\em iii})\xspace}

\makeatletter
\renewcommand{\paragraph}[1]{\vspace*{0.03in}\noindent{\bf #1.}\hspace{0.25ex \@plus1ex \@minus.2ex}}
\makeatother

\makeatletter
\newcommand{\paragraphc}[1]{\vspace*{0.03in}\noindent{\bf #1}\hspace{1ex \@minus.2ex}}
\makeatother
 \IEEEoverridecommandlockouts

\setacronymstyle{long-short}

\newacronym
  {adc}
  {ADC}
  {analogue-to-digital converter}

\newacronym
  {bti}
  {BTI}
  {bias temperature instability}

\newacronym
  {cots}
  {COTS}
  {commercial-off-the-shelf}

\newacronym
  {dl}
  {DL}
  {deep learning}

\newacronym
  {ewma}
  {EWMA}
  {exponentially weighted moving average}

\newacronym
  {fpga}
  {FPGA}
  {field-programmable gate array}

\newacronym
  {gaafet}
  {gate-all-around FET}
  {GAAFET}

\newacronym
  {gru}
  {GRU}
  {gated recurrent units}

\newacronym
  {hci}
  {HCI}
  {hot-carrier injection}

\newacronym
  {ic}
  {IC}
  {integrated circuit}

\newacronym
  {llrf}
  {LLRF}
  {low-level radio frequency}

\newacronym
  {loess}
  {LOESS}
  {locally estimated scatterplot smoothing}

\newacronym
  {lstm}
  {LSTM}
  {long short-term memory}

\newacronym
  {mape}
  {MAPE}
  {mean absolute percentage error}

\newacronym
  {ml}
  {ML}
  {machine learning}

\newacronym
  {pcie}
  {PCIe}
  {peripheral component interconnect express}

\newacronym
  {pdm}
  {PDM}
  {propagation delay measurement}

\newacronym
  {rf}
  {RF}
  {radio frequency}

\newacronym
  {rnn}
  {RNN}
  {recurrent neural network}

\newacronym
  {ro}
  {RO}
  {ring oscillator}

\newacronym
  {stl}
  {STL}
  {seasonal-trend decomposition procedure based on LOESS}

\newacronym
  {tddb}
  {TDDB}
  {time-dependent dielectric breakdown}

\newacronym
  {tid}
  {TID}
  {total ionizing dose}

\newacronym
  {xfel}
  {Eu-XFEL}
  {European X-ray Free-Electron Laser} 
\begin{document}

\bstctlcite{IEEEexample:BSTcontrol}

\title{Switching Frequency as FPGA Monitor: Studying  Degradation and Ageing Prognosis at Large Scale}

\ifanonymous
\author{\IEEEauthorblockN{Paper \#NNN, \pageref{lastpage}~pages (incl. references)}}
\else
\author{
    \IEEEauthorblockN{Leandro Lanzieri\IEEEauthorrefmark{1}\IEEEauthorrefmark{2}\IEEEauthorrefmark{3},
                      Lukasz Butkowski\IEEEauthorrefmark{1},
                      Jiri Kral\IEEEauthorrefmark{1},
                      Goerschwin Fey\IEEEauthorrefmark{2},
                      Holger Schlarb\IEEEauthorrefmark{1}, and
                      Thomas C. Schmidt\IEEEauthorrefmark{3} \\
    }

    \IEEEauthorblockA{
        \IEEEauthorrefmark{1}Deutsches Elektronen-Synchrotron DESY, Germany \\
        \{leandro.lanzieri, lukasz.butkowski, jiri.kral, holger.schlarb\}@desy.de
    }

    \IEEEauthorblockA{
        \IEEEauthorrefmark{2}Hamburg University of Technology, Germany $\cdot$ goerschwin.fey@tuhh.de
    }

    \IEEEauthorblockA{
        \IEEEauthorrefmark{3}Hamburg University of Applied Sciences, Germany $\cdot$ t.schmidt@haw-hamburg.de
    }
}
\fi

\maketitle

\begin{abstract}
The growing deployment of unhardened embedded devices in critical systems demands the monitoring  of hardware ageing as part of predictive maintenance.
In this paper, we study degradation on a large deployment of 298 naturally aged FPGAs operating in the European XFEL particle accelerator.
We base our statistical analyses on 280 days of in-field measurements and find a generalized and continuous degradation of the switching frequency across all devices with a median value of 0.064\%.
The large scale of this study allows us to localize areas of the deployed FPGAs that are highly impacted by degradation.
Moreover, by training machine learning models on the collected data, we are able to forecast future trends of frequency degradation with horizons of 60 days and relative errors as little as 0.002\% over an evaluation period of 100 days.
\end{abstract}


\section{Introduction}\label{sec:introduction}

Embedded devices have become integral parts of dependable systems and critical infrastructure, including distributed industrial and scientific facilities \cite{llrf-baag-12}.
While \gls{cots} components remain prevalent across critical deployments due to their reduced costs, their unhardened design exposes them to degradation mechanisms under adverse operational conditions.
Harsh environments exacerbate the effects of hardware ageing such as \gls{bti} and \gls{hci}, altering parameters of the transistors that compose embedded devices \cite{nbti-s-07, rahd-swz-10}.
These mechanisms progressively modify the characteristics of the underlying hardware, compromising the functionality and reliability of components, which can pose high risks with dangerous consequences.

\Glspl{fpga} are frequently deployed owing to their high speeds, low costs, and flexibility.
Nonetheless, degradation mechanisms affect \glspl{fpga}, reducing the propagation speed of signals over time \cite{lbkfs-sdpdf-24}.
This deterioration is particularly problematic on high-speed applications as it can negatively impact combinational circuits by modifying their critical paths and inducing timing violations.

\begin{figure}[t]
    \centering
    \includegraphics[width=\linewidth]{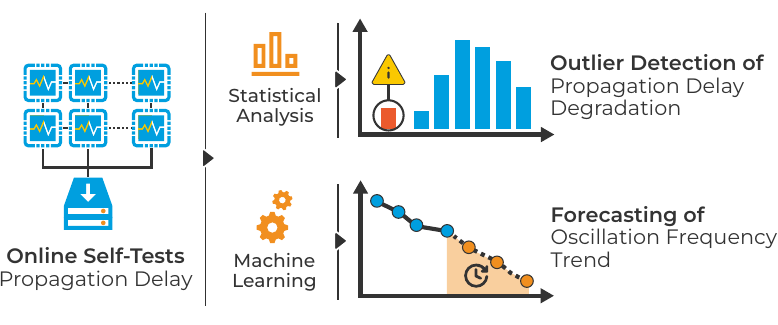}
    \caption{Propagation delay data from serves statistical analyses and machine learning models to evaluate the hardware degradation, detect outlier behaviour, and forecast frequency trends.}
    \vspace{-1pt}
    \label{fig:concept}
\end{figure}

Current research efforts actively focus on detecting degradation of \glspl{fpga} by utilizing \gls{ro} sensors, but most of the work in the literature evaluates hardware ageing either on accelerated or laboratory conditions \cite{lmfss-rtadm-24}.
Moreover, most of the related work studies a handful of devices at a time, possibly diminishing the statistical significance of the results \cite{shfs-zh-10, lmba-kknfi-23}.
This manifests the need for long-term and large-scale evaluations of the impact of degradation mechanisms on \glspl{fpga} in real-world deployments.

Predictive maintenance based on continuous monitoring and degradation forecasting surfaces as a crucial requirement to ensure dependability of mission-critical embedded systems that face hardware ageing.
Nevertheless, forecasting of degradation indicators on \glspl{fpga} has only been explored for smaller samples of artificially aged devices, which limits the generalization of these forecasts \cite{amem-sm-23,dmmua-sm-24}.

In this work, we bridge both research gaps by deploying online concurrent self-tests and extending our previous work \cite{lbkfs-sdpdf-24} on \num{298}~\glspl{fpga} exposed to a prolonged natural ageing process in operation at the \gls{xfel} particle accelerator \cite{xfel-06}.
In detail, our contributions are as follows:
\one we perform a study of the degradation in propagation delay across the devices using measurements taken during a period of \num{280}~days of operation;
\two we utilize time series analysis techniques to characterize degradation effects on the monitored devices over time;
\three we apply time series forecasting to estimate future trends of \gls{ro} frequency using \gls{ml} models and monitoring data.

Our time series study based on long-term oscillator measurements shows a steady degradation across all devices with a median shift in frequency of \qty{0.064}{\percent}, while our large-scale statistical analysis reveals highly degraded areas of the \gls{fpga} fabric.
Utilizing in-field monitoring data, we apply a backtesting evaluation schema, through which we are able to estimate future \gls{ro} frequencies with forecast horizons of \num{60}~days and relative errors of \qty{0.002}{\percent}.

The structure of this manuscript is as follows.
\autoref{sec:background} lays the background on relevant hardware degradation mechanisms and discusses related work.
\autoref{sec:method} describes the devices under study, the implementation of the self-tests, and the measurement procedures.
\autoref{sec:shutdown} and \autoref{sec:continuous} show and analyse the propagation delay data collected during shutdown and operation periods respectively.
\autoref{sec:forecasting} evaluates the performance of various techniques for time series forecasting on the collected data, and analyses their applicability as predictive maintenance systems.
The work concludes in \autoref{sec:conclusion} with an outlook.
 \section{Background and Related Work} \label{sec:background}

\subsection{Ageing Mechanisms in Embedded Devices}

Embedded systems are composed of components that undergo physical degradation during deployment.
Continuous wear-out processes exacerbated by environmental conditions modify the underlying hardware of the devices, which can result in performance reduction and failures.
In addition, trends in footprint reduction of silicon nodes increase the challenges originating from the ageing mechanisms at play in the devices.

\Gls{bti} is a degradation process that occurs in transistors, through which positive ions (\(H^+\)) diffuse and accumulate in the oxide of the gate, opposing the electric field generated by the gate-source activation voltage \cite{gdrm-os-95}.
As a consequence, the threshold voltage increases and the transistor transconductance degrades, which in turn can reduce the maximum switching frequency of the device.
The effects of this mechanism depend on the time the device remains active (\ie duty cycle), as well as the temperature.
Although this issue was initially observed on large nodes, \gls{bti} remains under active research due to its effect persisting on more modern technologies, such as FinFETs \cite{canv-pacm-19} and \glspl{gaafet}.

\Gls{hci} is an ageing mechanism that affects physical characteristics of transistors and depends on the switching activity of the device \cite{chcd-clcc-21}.
As a result of high electric fields applied to the devices, carriers gain enough kinetic energy to surpass the potential barrier of the gate oxide, diffusing into the dielectric and damaging the substrate interface.
Similarly to \gls{bti}, this process has the consequence of incrementing the threshold voltage of the device, which can permanently impact the switching frequency of the hardware.

\subsection{Degradation Sensors Based on Ring Oscillators}

\begin{figure}[t]
    \centering
    \includegraphics[width=\linewidth]{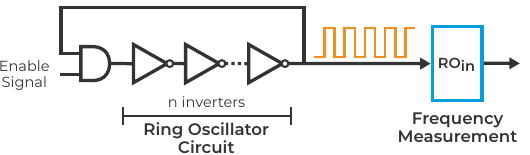}
    \caption{Propagation delay sensor based on a ring oscillator.}
    \label{fig:ro-sensor}
\end{figure}

\Glspl{ro} are popular building blocks for sensors that detect and monitor degradation of digital circuits, particularly \glspl{fpga} \cite{lmfss-rtadm-24}.
Although literature presents variations, a sensor is typically composed of the \gls{ro}, a circuit to measure the output frequency, and an optional enable input to switch oscillations on and off, as depicted in \autoref{fig:ro-sensor}.
The basic \gls{ro} circuit is constructed with an odd number of inverter gates connected in a way that they create a closed loop and form an astable circuit with an oscillating digital output.
At a given temperature and voltage, the frequency of the output signal \(f\) is given by

\begin{equation}
    f = \frac{1}{2 \cdot n \cdot t_p}
    \label{eq:ro_freq}
\end{equation}

\noindent where \(n\) is the number of inverters in the \gls{ro}, and \(t_p\) their propagation time of the underlying hardware.

This type of sensors are widely explored and employed in the literature.
Lu \etal describe a \gls{bti} monitor that was included on IBM z196 microprocessors \cite{bmlt-lj-13}.
Data of four systems over a period of \num{500}~days shows a noticeable degradation in their oscillation frequencies and an environmental dependency.
In turn, Kim \etal utilize two \glspl{ro} per sensor and measure the beat frequency to achieve higher sensing resolution \cite{soorm-kpk-08}.
With the goal of detecting recycled \gls{fpga} chips, Dogan \etal propose applying \gls{ml} classifiers on data of \gls{ro} degradation from \num{12} artificially aged \glspl{fpga} \cite{aarf-dft-14}.

Ageing monitoring and prediction based on \glspl{ro} is under active research on \gls{fpga} devices due to the flexibility of such sensors.
Li \etal studied hardware degradation on \num{24} artificially aged devices of \qty{28}{\nano\meter} for a theoretical equivalent usage time of \num{10}~years \cite{iama-lhwwl-22}.
Based on the laboratory measurements, the authors trained various \gls{ml} algorithms to predict degradation levels over time, which yielded a root mean squared error of \qty{0.292}{\percent}.
Furthermore, Zick \etal developed a footprint-optimized sensor for delay variations, power usage estimation and temperature \cite{shfs-zh-10}.
The authors observed a degradation on the propagation delay of two \glspl{fpga} placed under accelerated ageing conditions.
Moreover, Pfeifer \etal developed the concept of \enquote{reliability lab-on-chip} to perform calculations of propagation delays based on streams of signals \cite{rloc-pkp-14}.
The \gls{ro}-based system was evaluated on various \gls{fpga} devices, but only for a period of seven days.

Thanks to their relatively small resource occupation and ease of instantiation, \glspl{ro} are good candidates for continuous in-field measurements, and are often used in combination with \gls{ml} estimation models.
Kobayashi \etal evaluated various sensor architectures on a \qty{7}{\nano\meter} chip during a period of \num{144}~days \cite{lmba-kknfi-23}.
With the collected data, a simulation model was developed, which estimates \gls{bti} degradation and variation of threshold voltage based on traps.
Furthermore, Arvin \etal utilized neural network predictions of future \gls{ro} frequency values to detect anomalous behaviour due to the inclusion of trojan hardware on \glspl{ic} \cite{apic-aj-21}.
The authors were able to detect the presence of trojan circuits due to an increase in the average \qty{25}{\percent} forecasting error, caused by the extra activity in the vicinity of the \gls{ro}.
Sobas \etal also proposed a propagation delay estimation model after conducting a study of \num{8000}~hours \cite{amem-sm-23,dmmua-sm-24} to measure the degradation of \qty{16}{\nano\meter} FinFET \glspl{fpga} with \glspl{ro}.
This study reported errors of predictions in the order of \qty{10}{\percent}.

In previous work \cite{lbkfs-sdpdf-24}, we introduced a measurement module based on \glspl{ro} that allowed us to perform an initial large-scale set of measurements of \num{298}~boards of the \gls{llrf} system of the \gls{xfel}.
By comparing measurement results of operational and in-storage devices we found a significant degradation effect due to ageing mechanisms acting on the deployed devices.
In this work we extend our previous analysis based on new long-term measurements, which allow us to study the degradation behaviour over time and by location on the \glspl{fpga}.
Furthermore, we train and compare the performance of multiple \gls{ml} models on data collected during \num{280}~days from this large-scale deployment.
 \section{Measurement Methodology}\label{sec:method}

\subsection{Devices Under Study}\label{subsec:dut}

The \gls{llrf} system is a highly dependable and crucial part of the \gls{xfel} particle accelerator.
This system is in charge of precisely controlling the \gls{rf} signals that are applied on the cavities in order to correctly accelerate the electrons through the machine.
The system is distributed in \num{50} stations along kilometres of underground tunnel, and consists of multiple devices deployed on MicroTCA crates.
The crates provide device interconnection, power, mechanical support, and refrigeration to the boards that comprise the system.

\begin{figure}[t]
    \centering
    \includegraphics[width=\linewidth]{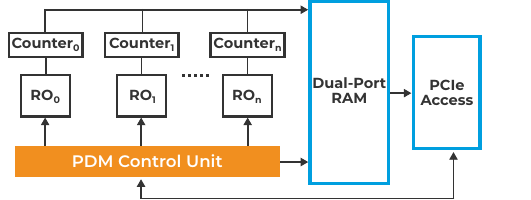}
    \caption{Propagation delay measurement module consisting of ring oscillators and counters managed by a control unit via the PCIe interface.}
    \label{fig:pdm_module}
\end{figure}

We base our study of propagation delay on data collected from \num{298} \glspl{fpga}, which are currently deployed at the accelerator.
More precisely, the devices under evaluation are Xilinx Virtex-6 XC6VLX130T \glspl{fpga} with speed grade -2 and built in \qty{40}{\nano\meter} technology.
These devices are part of digitizer boards in charge of performing measurements on the \gls{rf} cavities, distributed in groups of six devices per crate.
The hardware has been deployed and in operation in the accelerator facility for seven years.
During this period, all \glspl{fpga} have shared the same operational firmware.

\subsection{Propagation Delay Measurement Module}

Fault detection and predictive maintenance for systems such as the \gls{llrf} require continuous data collection originating from a large variety of heterogeneous \gls{fpga}-based devices.
To enable the collection of such data during operation, we developed a generic module that allows remote online and concurrent measurements of propagation delay \cite{lbkfs-sdpdf-24}.
The module is based on the open source DESY \gls{fpga} Framework \cite{fwk-bbbd-23}, which provides effortless integration into other platforms of the facility, as well as remote control via a common \gls{pcie} interface.

\autoref{fig:pdm_module} depicts the three main parts of the \gls{pdm} module: \one \gls{ro}-based sensors, \two control unit to locally manage the sensor activation, and \three unit to allow access via \gls{pcie}.
At design time, the module provides designers with control over the number of stages on the \glspl{ro}, as well as their location.
The control unit ensures that sensors are activated one by one for a pre-configured amount of time upon a measurement trigger.
Intermediate results are stored in RAMs exposed via \gls{pcie} registers to the user, and are collected in parallel by a measurement software.

\subsection{Measuring Propagation Delay on FPGAs}

In this study, we analyse two types of measurements of the propagation delay: at shutdown and continuous.
Both types are performed on the same devices but under different operating conditions.
For these measurements, we configure the \gls{pdm} to distribute the \glspl{ro} and fix their locations on the \gls{fpga} fabric.

A measurement software is in charge of the accesses to all devices simultaneously during the data collection period and triggering the \gls{pdm}, which in turn controls the process locally.
The local control unit ensures that every \gls{ro} oscillates during \qty{500}{\milli\second} at a time while the counter registers the amount of oscillations.

\paragraphc{Shutdown measurements} are performed twice during facility maintenance periods, six months apart from each other.
As target devices are not in operation during shutdowns, measurements do not require sharing \gls{fpga} resources with the operative application.
Therefore, the measurement firmware is composed of \num{100} \glspl{ro} distributed over the device, in addition to the necessary logic to control the oscillators and access the \gls{pcie} registers.
This amount of oscillators provides a good spatial resolution for the evaluation of propagation delay over long periods of time.
This measurement setup totals \num{29800} \glspl{ro} across all deployed \glspl{fpga}, each of which is measured \num{100}~times.
It is worth noting that we employ the exact same firmware on both shutdown measurements.

\paragraphc{Continuous monitoring} operates during service time owing to the flexibility of the \gls{pdm} module, which allows integrating it into the functional application of the devices under study.
This way, measurements can be executed with devices on normal service, allowing for a constant collection of data during the accelerator operation.
Placing a self-test unit alongside the application means that both must share the \gls{fpga} resources.
For that reason, the continuous monitoring module deploys \num{8} \glspl{ro} per device, which provides around \num{2380} \glspl{ro} over the entire group of \glspl{fpga}.
Although this setup has less spatial resolution in comparison to the shutdown measurements, it allows us to perform measurements much more frequently, which provides a much more detailed understanding of the time dependence.
Therefore, data collected from continuous monitoring is conformed of measurements taken every \num{2}~hours during \num{280}~days.

In addition to the \gls{ro} frequency data, environmental and device covariates are collected during the tests.
The measurements record sensors from the \gls{fpga} boards, such as core voltage and temperature, as well as associated devices, including power supplies and cooling units.
\autoref{tab:covariates} comprehensively lists and describes the collected environmental covariates.

\begin{table}[t]
    \caption{Environmental covariates collected during ring oscillator measurements.}
    \label{tab:covariates}
    \centering
    \begin{tabular}{|c|l|}
    \hline
    \multicolumn{1}{|c|}{\bfseries Magnitude} & \multicolumn{1}{c|}{\bfseries Covariate} \\
    \hline \hline
    \multirow{6}{*}{\begin{minipage}{5em} \centering Voltage \end{minipage}} & Core \qty{1.0}{\volt} rail \\
    \cline{2-2}
    & DDR \qty{1.5}{\volt} rail \\
    \cline{2-2}
    & \qty{1.8}{\volt} rail \\
    \cline{2-2}
    & \qty{2.5}{\volt} rail \\
    \cline{2-2}
    & \qty{12}{\volt} payload rail \\
    \cline{2-2}
    & \qty{3.3}{\volt} management rail \\
    \cline{1-2}
    \multirow{5}{*}{\begin{minipage}{5em} \centering Temperature \end{minipage}} & FPGA die \\
    \cline{2-2}
    & Board \\
    \cline{2-2}
    & Refrigeration inlet \\
    \cline{2-2}
    & Refrigeration inner \\
    \cline{2-2}
    & Refrigeration outlet \\
    \cline{1-2}
    \multirow{2}{*}{\begin{minipage}{5em} \centering Current \end{minipage}} & MicroTCA RTM management \\
    \cline{2-2}
    & MicroTCA RTM payload \\
    \hline
    \end{tabular}
\end{table} \section{Propagation Delay Between Shutdowns}\label{sec:shutdown}

On shutdown periods of the \gls{xfel}, the \gls{fpga} boards under study remain installed but are out of operation.
During this maintenance time, we have the opportunity of performing detailed measurements of the propagation delay \textit{in situ}, by utilizing a custom firmware that does not disturb the application.
In this section, we compare and analyse measurements taken at two consecutive shutdown periods, six months apart.

\subsection{Shutdown Measurements}

To understand the variation of propagation delay between measurement points, we calculate the shift \(\Delta^d(i)\) of the median oscillation frequency for each \gls{ro} \(i\) of a device \(d\), relative to the frequency of the initial measurement, as shown in \autoref{eq:freq_shift}.
\(\widetilde{f^d_0}(i)\) and \(\widetilde{f^d_1}(i)\) are the median frequencies on the initial and final measurement respectively.

\begin{equation}
    \Delta^d(i) = \frac{\widetilde{f^d_1}(i) - \widetilde{f^d_0}(i)} {\widetilde{f^d_0}(i)}
    \label{eq:freq_shift}
\end{equation}

\begin{figure}[t]
    \centering
    \includegraphics[width=\linewidth]{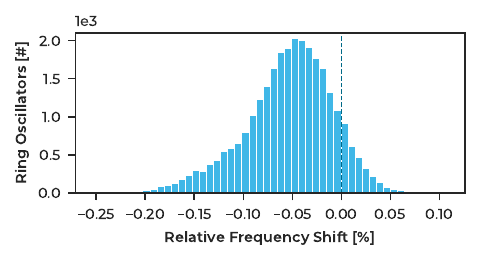}
    \caption{Distribution of relative frequency shift of the oscillators with a median degradation of \qty{-0.0496}{\percent} over a period of \num{6}~months between shutdowns.}
    \label{fig:hist_shift}
\end{figure}

\autoref{fig:hist_shift} shows that more than \qty{90}{\percent} of oscillators experimented a negative shift, meaning a slower frequency after the six-months operation period.
Interestingly, a small share of the \glspl{ro} presents positive shifts, indicating a possible effect of degradation recovery thanks to removed stress.
This can be due to changes in the operational firmware introduced after the initial shutdown measurements, which modified the locations of logic cores.
Indeed, previously stressed areas can undergo a slight recovery phase from effects like \gls{bti} when the stress is reduced.
Nonetheless, the negatively skewed distribution of shifts has a median of \qty{-0.0496}{\percent} and extremely negative values reaching further than \qty{-0.25}{\percent}.
This clearly indicates that most of the devices present increased propagation delays due to degradation suffered in operation.

\subsection{Anomaly Detection of Degraded Devices}

Significantly increased degradation of the propagation delay may be a performance issue on certain applications where strict timing constraints must be met.
Additionally, outlying degradations may be early indications of potential problems on the devices or their installations.
By periodically monitoring the overall ageing of the devices, early alerts can be issued to flag hardware that presents abnormal degradation.
Anomaly detection algorithms can be employed to automatically supervise the results of the periodic shutdown measurements and report upon abnormal behaviour.

The Z-score (also known as standard score) if often employed to detect outliers on datasets by observing how distinct a sample is from the data mean.
This metric normalizes the values, and is calculated as a relative deviation of each sample from the mean value of the measurements.
The issue with the traditional Z-score calculation is that it utilizes non-resistant data summaries (mean and standard deviation), which have a low breakdown point.
As a consequence, outliers can have a significant impact on the calculation of the score itself.

To determine which devices display a significantly different reduction in the oscillation frequency, we employ a modified Z-score \cite{hdho-ih-93}, which relies on a resistant data summary: the median value.
For each device \(d\), we calculate a modified Z-score \(Z^d\) as

\begin{equation*}
    Z^d = \frac{0.6745 \cdot (\widetilde{\Delta^d} - \widetilde{\Delta})} {\text{MAD}}
    \label{eq:modified_z_score}
\end{equation*}

\noindent where \(\widetilde{\Delta^d}\) is the median frequency shift of the \glspl{ro} of device \(d\), \(\widetilde{\Delta}\) is the median shift across all devices. MAD stands for median absolute deviation, and can be calculated as follows

\begin{equation*}
    \text{MAD} = \text{median}(\vert \widetilde{\Delta^d} - \widetilde{\Delta} \vert)
    \label{eq:mad}
\end{equation*}

\begin{figure}[t]
    \centering
    \includegraphics[width=\linewidth]{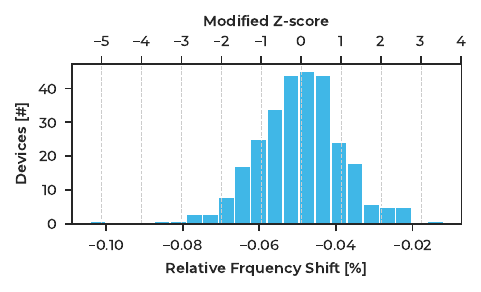}
    \caption{Distribution of median relative frequency shift with their corresponding modified Z-scores, aggregated per device between shutdown measurements.}
    \label{fig:hist_difference_dev_agg}
\end{figure}

A value of \num{+-3.5} is commonly considered as a threshold \cite{hdho-ih-93} above which data points are flagged as outliers of the dataset.
This value can be adjusted to tune the sensibility of the anomaly detection process.
As an alternative, multiple values can be applied to provide different levels of early alerts.

\begin{figure}[t]
    \centering
    \includegraphics[width=\linewidth]{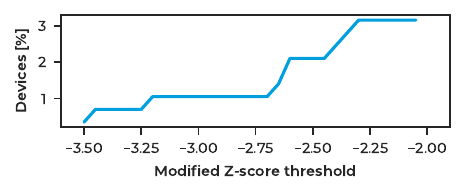}
    \caption{Share of detected outlying devices at different Z-score thresholds.}
    \label{fig:outliers}
\end{figure}

\autoref{fig:hist_difference_dev_agg} shows the distribution of median frequency shifts aggregated per device, together with the corresponding modified Z-scores.
We can observe that all devices display a shift below zero, with the median located at \qty{-0.05}{\percent} and values as negative as \qty{-0.10}{\percent}.
Analysing the median shift per device masks the effect of \glspl{ro} with positive frequency changes, and clearly shows that all devices tend towards slower propagation delays after the six months of operation.
Additionally, the figure depicts a clear device with significantly negative Z-score and frequency shift.
\autoref{fig:outliers} summarizes the share of devices found by the outlier detection at different threshold values.

As an additional test to the anomaly detection, we evaluate the modified Z-score on a replacement device of the \gls{llrf} system.
The replacement device, which was previously in operation at a different facility, has been installed after the initial shutdown measurements and before the final shutdown measurements.
The measurement yields a median frequency shift of \qty{-11.24}{\percent} between both devices.
This shift corresponds to a Z-score of \num{-1072}, which is largely outside the threshold and is clearly detected as an anomalous variation.

\subsection{Physical Distribution of the Frequency Shift}

A distinctive advantage of \gls{ro}-based sensors is their locality, which can provide a particularly detailed spatial understanding of variations in the propagation delay within the \gls{fpga}.
This knowledge gives designers valuable insights into the effects that certain logic cores and their usage patterns have on the underlying \gls{fpga} hardware.
Insights are particularly useful for deployments where periodic firmware updates are possible, as the location information provided by measurements can be employed to change the placement of logic that has tight timing constraints to areas with less degradation.
In addition, healing strategies can be applied by rotating the placement of highly active cores so that degraded areas have recovery time.

\begin{figure}[t]
    \centering
    \includegraphics[width=\linewidth]{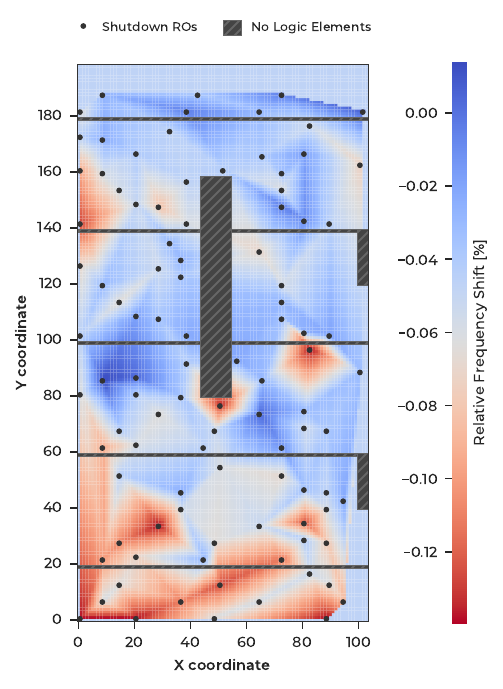}
    \caption{Distribution of median frequency shifts between shutdowns across FPGA devices, linearly interpolated between ring oscillator locations.}
    \label{fig:location_difference_median}
\end{figure}

With the information collected across the \num{100} distributed \glspl{ro} on the deployed \glspl{fpga}, we can perform an estimation of the propagation delay at various points of the programmable fabric of the devices.
By calculating the median delay across all boards and performing a linear interpolation between the measurement locations, we can build a representative degradation map, as shown in \autoref{fig:location_difference_median}.
The grey circles show the locations of the \glspl{ro} that take propagation delay measurements during shutdowns.
Areas in shades of red represent more negative frequency shifts between shutdown measurements, which indicate a higher degree of degradation over time.
In turn, stronger blue regions show reduced and even positive frequency variations.
We can observe hotspots of degradation concentrated mostly in the lower region of the \glspl{fpga}, particularly on the left side.
To better understand the degradation effects of the circuits that are deployed during operation, we perform a correlation between the highly impacted areas and the logic that is placed with the operational firmware.

By manually inspecting the circuits placement, we find that DDR3 DRAM logic extensively occupies the regions in the lower left and right corners, as many pins on that side of the chip connect the external RAM.
This logic is not only constantly active during operation due to the continuous data acquisition, but also works at high frequencies communicating with the memory chip at \qty{400}{\mega\hertz}.
Such active usage pattern can increase temperatures and accelerate degradation processes such as \gls{bti} and \gls{hci} in the underlying hardware of the area.

The hardware under study operates as high-speed data acquisition and computing units, constantly taking in digital samples from external \glspl{adc} and performing signal processing tasks on them.
The firmware inspection reveals that most of the computing logic is placed in the area around the hotspot surrounding the \gls{ro} on coordinates \((X=82; Y=96)\).
This is a clearly degraded region in comparison to the surrounding areas.

This degradation analysis provides insights on the tangible effects that different logic cores and usage patterns cause on the hardware.
Including this information into design pipelines can aid developers in actively taking precautionary actions when updating firmware versions.

 \section{Continuous Monitoring Analysis}\label{sec:continuous}

In this section, we present the results of the continuous measurements performed on the \gls{fpga} devices.
We analyse the time series with methods to determine whether signs of degradation over time are observable in the deployment.

\begin{figure}[t]
    \centering
    \includegraphics[width=\linewidth]{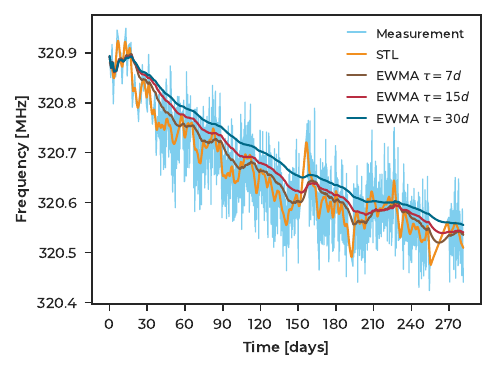}
    \caption{Comparison of trend extraction techniques on frequency measurements from a ring oscillator. Trends show a relatively steady degradation and negative mean slope over the \num{280}~days of monitoring.}
    \label{fig:example_ts}
\end{figure}

\subsection{Extraction of Measurements Trend}\label{subsec:trend}

In order to analyse how the propagation delay varies over time on the studied devices, we perform a statistical analysis of the measurement time series.
We focus on measurement trends, given that they reduce noise and hold information about the evolution of propagation delay over long periods of time.
We explore two methods to reduce measurement noise and to extract the trend from the data.

One technique to extract trends from time series is the \gls{stl} \cite{stl-ccmt-90}, where \acrshort{loess} stands for \acrlong{loess} \cite{loess-cd-88}.
\glsunset{loess}
\Gls{loess}, sometimes also referred to as locally weighted polynomial regression, is a modelling method that fits a low-degree polynomial model at each point of the dataset.
The fitting process applies weighted least squares, which gives more weight to points in the vicinity of the approximated value and less weight to points further away, similarly to \gls{ewma}.
Once the polynomial models are determined, the trend can be calculated by evaluating them at each point of the dataset.

\Gls{ewma} \cite{ewma-h-86} smoothing is a commonly used technique to pre-process data and to reduce noise.
The main idea behind it is to average the values within a moving window, where the weighing of previous samples decays exponentially with the distance to the evaluated point in time.
Therefore, samples further from the evaluated one have a reduced effect on the current value.
The smoothing parameter \(\alpha\) determines the weight of previous samples in the exponential decay.
This parameter can also be given in terms of the \emph{half-life} \(\tau\), which represents time lag at which the exponential weights decay by \num{0.5}, as shown in \autoref{eq:half_life}.

\begin{equation}
    \alpha = 1 - \exp(- \ln(2) / \tau) \quad ; \quad \tau > 0
    \label{eq:half_life}
\end{equation}

\autoref{fig:example_ts} shows measurements taken from a \gls{ro} during the \num{280}~days monitoring of the \gls{llrf} devices, together with the results of the trend-extraction approaches, including \gls{ewma} using various \(\tau\) settings.
The light blue line shows the raw measurements recorded on the device every \num{2}~hours, which displays a significant amount of noise.
We can observe that the \gls{stl} performs a certain degree of noise reduction, but it is still very much affected by local variations of the measurements.
As expected, the noise in the trends extracted with \gls{ewma} decreases as the \(\tau\) parameter increases because older values are more strongly weighted in the averaging process masking high frequency changes.
As the overall long-term trends of the frequencies are the principal object of study, we choose \gls{ewma} with a \(\tau\) of \num{30}~days for trend extraction on the remaining of the analysis.

Even though the filtered series are not monotonic and still display noise, it is easier to observe the trend in its values over time.
The downward trend on the data points is a consequence of the \gls{ro} increasing the propagation delay during operation time.
This shows that trend extraction aids in enhancing long-term and slower changes in the \gls{ro} frequency.

\begin{figure}[t]
    \centering
    \includegraphics[width=\linewidth]{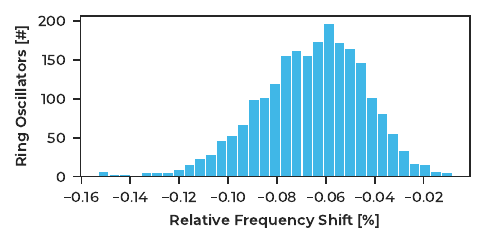}
    \caption{Distribution of the relative frequency shift between the first and the last \num{30}~days of the \num{280}-days monitoring period, with a median of \qty{-0.064}{\percent}.}
    \label{fig:distribution_timeseries}
\end{figure}

To achieve an estimation of the frequency variation, we compare the frequency trend values at the beginning and end of the monitoring period.
We take the median values over the initial and final \num{30}~days of measurements for each \gls{ro}, and we calculate the relative shift.
\autoref{fig:distribution_timeseries} depicts the distribution of these relative frequency shifts, all of which present negative values.
The median frequency shift is \qty{-0.064}{\percent}, which is slightly more negative than the median value on the \num{6}-months shutdown measurement (\cf \autoref{sec:shutdown}).
This suggests that all \gls{ro} present slower switching frequencies after the monitoring period of \num{280}~days.
To determine whether this possible downward frequency trend is a widespread behaviour over time across the studied devices, we perform a statistical analysis of the trend of the time series across all \glspl{ro}.
Given the time-dependent nature of the series, with this analysis we aim to obtain parameters that can encode information about the behaviour of the devices during the monitoring period.

\subsection{Linear Regression Analysis}\label{subsec:regression}

As a first approach to the time series analysis, we perform a regression by fitting a linear model to each of the series and observe the slope values.
The slope of the calculated approximation encodes information regarding the rate of change of the series.
To this end, we employ the linear regression implementation from the Scikit-learn library \cite{sklearn-pvgmt-11}.

\begin{figure*}[t]
    \centering
    \begin{subfigure}[t]{0.48\textwidth}
        \centering
        \includegraphics[width=\linewidth]{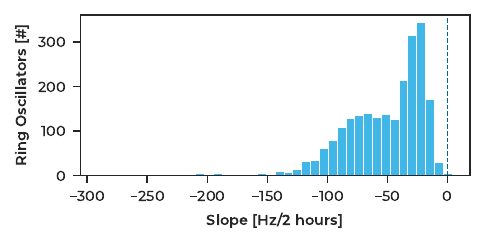}
        \caption{Slopes from linear regressions fitted to the time series trends.}
        \label{fig:slopes_distribution}
    \end{subfigure}
    \hfill
    \begin{subfigure}[t]{0.5\textwidth}
        \centering
        \includegraphics[width=\linewidth]{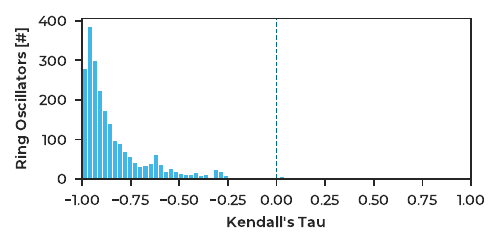}
        \caption{Kendall Tau statistic between frequency values and time axis.}
        \label{fig:taus_distribution}
    \end{subfigure}
    \caption{Result distributions from the statistical analyses of the frequency trends of the ring oscillators.}
    \label{fig:combined_distribution}
\end{figure*}

\autoref{fig:slopes_distribution} depicts a histogram with the distribution of the slopes extracted from the fitted linear regression models.
We can observe a negatively-skewed distribution, where more than \qty{97}{\percent} of the slopes have values below zero.
In fact, the median value lays at \num{-42.33}, while the $25^{th}$ and $75^{th}$ percentiles are \num{-74.33} and \num{-25.93} respectively.
This distribution strongly indicates a decreasing trend in time for most of the \glspl{ro} frequencies, while only few present upward trends or no strong trend at all.

\subsection{Correlation Statistical Test}\label{subsec:correlation}

A second method to analyse the trend of the time series derived from the measurements is to statistically evaluate the correlation between frequency and time.
To avoid the requirement of assuming a data distribution, we apply Kendall's rank correlation \cite{nmrc-k-38}, also known as Kendall's Tau.
This non-parametric statistic evaluates two ordered sequences and estimates their correlation based on the amount of concordant and discordant data pairs found between the sequences.
The possible values of Kendall's Tau range between \num{-1} and \num{1}.
Statistic results close to \num{1} or \num{-1} suggest a strong correlation or anti-correlation respectively, while results close to \num{0} suggest that no correlation exists.

The calculation of Kendall's Tau works as follows: consider two pairs \((i, Y_i)\) and \((j, Y_j)\), where \(i\) and \(j\) are points in time, and \(Y_i\) and \(Y_j\) are the frequency values at those points. The pairs are considered concordant if \(Y_i < Y_j\) and \(i < j\), and they are discordant when \(Y_i > Y_j\) and \(i < j\). As a consequence, Kendall's Tau is a statistic sensitive to the ordinal correlation between the sample and the time axis. The value can be calculated as

\begin{equation*}
    \tau = \frac{C - D}{C + D}
    \label{eq:kendall}
\end{equation*}

\noindent where \(\tau\) is the statistic, \(C\) represents the amount of data pairs that are concordant, and
\(D\) the amount that are discordant.

To properly evaluate how statistically significant the correlations indicated by the statistic values really are, hypothesis tests are required.
The objective of the test is to assess whether there is enough evidence in the data to confidently accept the calculated correlation.
This is done by evaluating the probability of obtaining a correlation value as extreme as the calculated one in a case where a no-correlation null hypothesis is true.
The considered null hypothesis \(H_0\) in this case is that frequency values of a \gls{ro} and time are independent, meaning that the expected value for \(\tau\) is zero.
In contrast, the alternative hypothesis states that there is indeed a correlation between time and frequency.

The hypothesis test is conducted by comparing the calculated \(\tau\) value against the null distribution, which is approximated as a normal one with a variance dependent on the number of samples in the case of Kendall's Tau.
The comparison yields a p-value, which represents the proportion of values in the null distribution that are as extreme or more than the calculated \(\tau\).
In other words, the p-value is the probability of observing the calculated \(\tau\) on the data given that \(H_0\) is true.

To reject the hypothesis \(H_0\) in favour of the alternative, a significance level \(\alpha\) is chosen and compared against the p-value.
A p-value lower than \(\alpha\) indicates that the null hypothesis is not likely, and the alternative should be favoured instead.
Therefore, \(\alpha\) represents the probability of rejecting \(H_0\), given that \(H_0\) is true.

For this analysis we select a significance level \(\alpha = 1\%\), which means that we only accept \(\tau\) values where the probability of a false correlation is less than \qty{1}{\percent}.
The SciPy implementation \cite{scipy-vgohr-20} of Kendall's Tau that we utilize yields a p-value as well.
Therefore, we obtain a p-value and a \(\tau\) for each evaluated \gls{ro} frequency time series.

To evaluate the entire population of \glspl{ro}, we calculate the statistic for all time series, and then filter out the cases where the computed correlation is not statistically significant (\ie \(\text{p-value} < \alpha\)), which represents only \qty{0.39}{\percent} of the series.
This yields the \(\tau\) distribution shown in \autoref{fig:taus_distribution}.
We can observe that the correlation values are distributed with a positive skew, and lean towards the negative limit of the statistic with a median value of \num{-0.90}.
Most of the values show a high anti-correlation between frequency and time \cite{m-gucc-12}, indicating a reduction over time.
There is a small amount of positive \(\tau\) values with negligible correlations, which could be due to the presence of certain \glspl{ro} with a slightly increasing frequency over time.
This is aligned with some positive slope values that were found in \autoref{subsec:regression}.

Based on the statistical analysis we can conclude that most of the monitored \glspl{ro} display a reduction of their oscillation frequency over time.
This behaviour reflects the degradation of the underlying hardware as a consequence of the ageing mechanisms at play on the deployed devices.
 \section{Forecasting Propagation Delay}\label{sec:forecasting}

Achieving predictive maintenance of the deployed hardware requires estimating the evolution of degradation mechanisms via the monitored parameters.
To this end, we study the performance of \gls{ml} models to forecast the oscillation frequency of \glspl{ro}, based on data collected during continuous monitoring.
We perform a comparative analysis across models and evaluate them on various configurations and forecast horizons.

\begin{figure*}[t]
    \centering
    \includegraphics[width=\linewidth]{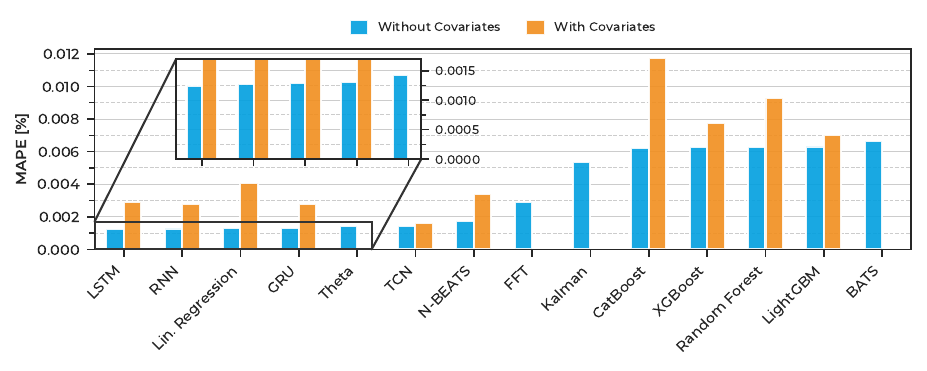}
    \caption{Mean average percentage errors yielded by various models when estimating the frequency trend of a ring oscillator over a forecast horizon of \num{70} days with and without covariates.}
    \label{fig:evaluation}
\end{figure*}

\subsection{Time series Forecasting}

Monitoring data is typically collected, stored, and studied as time series that represent data points over time.
The analysis of time series allows the extraction of time-related features, such as trend or seasonality, which in turn can be used to guide decisions.
The study monitoring data in the time domain adds a new dimension to anomaly detection, with which one can flag anomalous behaviour based on time patterns.

The forecast of time series estimates future values and trends of the studied variables based on historical data.
This can be done by extracting patterns from previous data and extrapolating them beyond the train dataset up to a forecast horizon.
Time series can be univariate, where a single variable is tracked over time, or multivariate, in case multiple variables are sampled at each point.

There is a variety of models that forecast time series with different capabilities.
Certain models work only with univariate series, while others allow multivariate forecasting.
Many models provide probabilistic forecasts, as they yield a probability distribution per time step instead of a single value.
In many cases, time series of additional variables encode information of the forecasted variable.
These additional variables are known as covariates, and certain forecasting models include them as a part of the training data to provide a context for the target variable.

\subsection{Comparative Evaluation of Models}\label{sec:evaluation}

Our goal is to determine which models and parameters perform the best in the estimation of future trends of propagation delays.
We train multiple traditional and \gls{dl} \gls{ml} models implemented with the Darts Python library \cite{darts-hlpsn-22}, which provides a homogeneous interface and utilities for time series manipulation.
The hyperparameters of each model are optimized via a random search with the Optuna optimization framework \cite{optuna-asyo-19}.

To compare the forecasting performance of the models, we train and evaluate them with data from the \gls{ro} depicted in \autoref{fig:example_ts}.
The frequency time series is pre-processed to extract the trend with the \gls{ewma} method configured with a \num{30}-days \textit{half-life}.
The \num{280}-days long data is split into two datasets: \num{210}~days (\qty{75}{\percent}) for training and \num{70}~days (\qty{25}{\percent}) for evaluation.
We employ the \gls{mape} as a common metric to compare the performance of the models, given that the value in percentage is conveniently interpretable.
The error is calculated as shown in \autoref{eq:mape}, where \(y\) is the actual value of the frequency trend, \(\hat{y}\) is the forecasted value, and \(V\) is the number of samples of the validation series.

\begin{equation}
    \text{MAPE}(y, \hat{y}) = \frac{100}{V} \sum_{t=1}^{V} \left\lvert \frac{y_t - \hat{y_t}}{y_t} \right\rvert
    \label{eq:mape}
\end{equation}

We evaluate the models under two configurations: \one without covariates, and \two with covariates.
In the first configuration, the model only receives the values of the \gls{ro} frequency trend as input.
In the second case, the environmental variables that have been recorded during operations (\cf \autoref{tab:covariates}) are included as covariates for training.

\autoref{fig:evaluation} summarizes the results of all evaluated models, sorted from left to right in increasing order of \gls{mape}.
Blue bars show errors for models trained only with \gls{ro} data, and orange bars represent errors when covariates are included in the training process.
Certain models do not allow including covariates, namely: theta, FFT, Kalman, and BATS.

The four best-performing models yield similar errors in the order of \qty[group-digits=none]{0.00125}{\percent}, and are closely followed by the theta model with roughly \qty{0.0015}{\percent}.
We observe that the errors are well below the median frequency change of \qty{0.064}{\percent} observed over the \num{280} monitoring days, as shown in \autoref{subsec:trend}.
This indicates that the models are good candidates for the estimation of frequency degradation during operation, which can be employed for predictive maintenance systems.

Three out of the four best models are different versions of the same type of networks: vanilla \gls{rnn}, \gls{lstm}, and \gls{gru}.
\Glspl{rnn} \cite{pdpa-j-97} are \gls{dl} models that contain recurrent connections (\ie cycles) with the goal of capturing sequence dynamics in the data.
By including previous values in the output computation, the model has information about past events which can then influence future outputs.
Although useful, simple \glspl{rnn} present an issue when multiple layers are included to capture long-term dependencies in the data: performing backpropagation becomes unfeasible due to vanishing and exploding gradients \cite{lldd-bsf-94}.

\Gls{lstm} models \cite{lstm-hs-97} alleviate this issue by including a new intermediate storage unit known as memory cell.
This cell keeps an internal state, which can selectively influence the network hidden state and output.
Multiple gates control the internal state: input gate, forget gate, and output gate.
These gates provide the model with mechanisms to update or reset the internal state depending on the input and the hidden state.
\Glspl{gru} \cite{opnml-cvbb-14} were developed as streamlined optimization based on the idea of \glspl{lstm}.
By only containing a reset gate and an update gate to modify the hidden state of the memory cell, \glspl{gru} achieve similar performances with lighter computation.

The linear regression model utilizes lag features, which are observations from the data shifted in time.
When additional covariates are included, they are considered input features for the linear combination as well.
This model fits multiple weights to the employed features with the least squares method.

Closely behind and in the fifth position is the theta model \cite{ttm-an-00}, which utilizes a decomposition approach to forecast time series by fitting two theta-lines to the data.
These lines change the local curvatures of the time series while keeping the mean and slope values, which approximates long-term behaviours of the data or augments short-term features depending on \(\theta\).

Interestingly, the inclusion of environmental variables shows a negative impact in performances: all evaluated models yield larger errors when trained with covariates.
This suggests that instantaneous environmental data does not have a strong influence in the long-term frequency trend, but rather contributes to the measurements noise.
Therefore, the addition of these variables does not provide extra information when forecasting \gls{ro} frequency trends over long periods of time.

\subsection{Backtesting for Predictive Maintenance}

\begin{figure}[t]
    \centering
    \includegraphics[width=\linewidth]{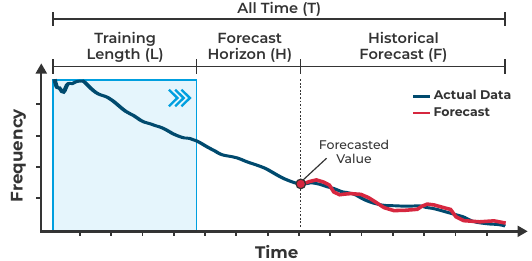}
    \caption{Backtest iteratively trains models on historical data to evaluate their performance as if the had been deployed in the past.}
    \label{fig:backtesting}
\end{figure}

To analyse the suitability of the models for predictive maintenance applications, we further evaluate the way they would have performed historically on the available \gls{ro} data, had they been deployed in the past.
We apply a test technique known as backtesting, where multiple training sets are built from historical data over all available time (\(T\)) by rolling a window of constant training length (\(L\)) stepwise, as illustrated in \autoref{fig:backtesting}.
Each data point produces a slightly different training set, on which we fit the models.
On each iteration, models yield a forecast at the end of the forecast horizon (\(H\)) time period.
By iteratively fitting and forecasting over all available training sets, the evaluated models produce historical forecasts with a length \(F\), given by \autoref{eq:hist_forecast}.
Finally, the \gls{mape} is calculated between the forecasted time series and the real data to obtain an estimation of how the models would have performed historically.

\begin{equation}
    F = T - (L + H)
    \label{eq:hist_forecast}
\end{equation}

In terms of predictive maintenance, the trade-off between forecast error and forecast horizon should be evaluated.
Forecasting values longer into the future can affect a model performance, but for certain cases an increment in the error can be tolerated for the valuable foresight.
Additionally, the inclusion or omission of past values can influence forecasting errors, therefore it is important to evaluate how the length of the training dataset modifies the model results at different forecast horizons.

As the length of the historical forecast \(F\) depends on \(L\) and \(H\), backtesting with different configurations yields time series of different lengths.
Therefore, special care should be taken when evaluating the \gls{mape} across configurations.
To fairly compare errors, we calculate them only over a constant period of time, which is the length of the shortest historical forecast yielded by the backtest procedure.
In our case, the shortest time series produced is \num{100}~days long, which corresponds to the configuration of \(L = 120\) and \(H = 60\) days.
This time represents more than \qty{36}{\percent} of the available data \(T\).

By utilizing the backtesting technique configured at different values of training length and forecast horizon, we evaluate two of the best-performing model presented in \autoref{sec:evaluation}: linear regression and theta.
We choose to backtest these models due to the shorter training time required in comparison to the \gls{rnn}-based models, and the similar errors in the comparative evaluation.
Such characteristics are desired in models deployed for predictive maintenance, where a continuous re-training with new data is common.

\begin{figure}[t]
    \centering
    \includegraphics[width=\linewidth]{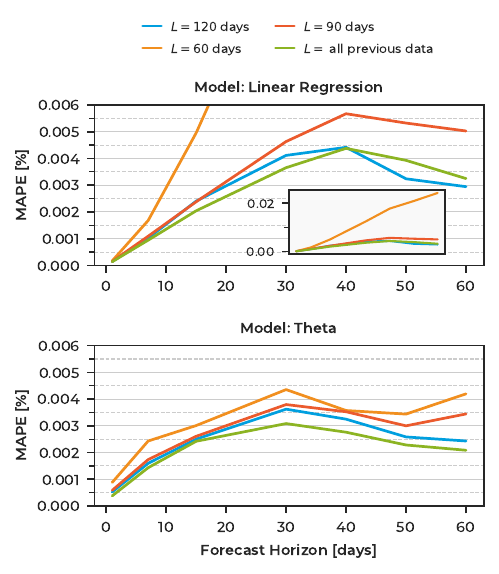}
    \caption{Backtest results of linear regression and theta models evaluated on various forecast horizons and training lengths (\(L\)).}
    \label{fig:backtest_results}
\end{figure}

\autoref{fig:backtest_results} shows the results of backtests evaluated on horizons between \num{1} and \num{60}~days, and with training lengths of \num{60}, \num{90}, and \num{120}~days.
We also include the case where the length of the training window increases on each iteration, so that all previous data is used to fit the model.
Each line colour represents the \gls{mape} of the model trained with a given training length (\(L\)).

The linear regression displays an almost linear relation between error and forecast horizon length until \num{40}~days, but with different slopes depending on the training length \(L\).
The case of \(60\)~days is an outlier with relative errors raising beyond \qty{0.02}{\percent}.
This model shows a strong dependence on the selection of the training window parameter, particularly for longer horizon days.

For the theta model, we also observe that increasing forecast horizons tends to increment prediction errors non-linearly for all training lengths, but only until \num{30}~days.
The model can forecast frequency trends with errors in the order of \qty{0.002}{\percent} even when predicting values \num{60}~days into the future.
The performance dependency of theta on training lengths is less pronounced than with the linear regression, as all configurations show a similar behaviour and the spread is smaller.
With forecast horizons greater than one month, the error has a local minimum at around \num{50}~days, after which the behaviour depends on the training length: there is an upward trend for \(L = 60\) and \(L = 90\), while the error decreases slightly for longer training windows.

Both model backtests show agreement in that longer training lengths (\ie larger training datasets) lead to lower errors for the same forecast horizon.
Nonetheless, the theta model presents a much more uniform behaviour for different configurations and an overall better performance even for a forecast horizon of two months.
From this comparison, it is clear that choosing too small training lengths can reduce the model performance, and that selecting fitting values for specific use cases is important.

To explore the forecast generalization of the theta model under the current configuration, we apply the same backtesting technique over the entire dataset of measurements from all \glspl{ro} and devices.
For this evaluation, we select the configuration with a forecast horizon of \num{60}~days and the best-performing training window of variable length, which utilizes all available previous data.
This backtest setup produces historical forecasts of \num{180}~days, over which the \gls{mape} is calculated for each \gls{ro} time series.

\begin{figure}[t]
    \centering
    \includegraphics[width=\linewidth]{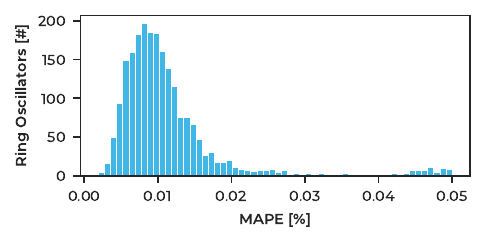}
    \caption{Distribution of mean average percentage errors yielded when backtesting all ring oscillator measurements with the theta model.}
    \label{fig:distribution backtests}
\end{figure}

\autoref{fig:distribution backtests} depicts a histogram of the backtesting errors for all \glspl{ro}.
We observe a positive skew in the \gls{mape} distribution, a median value of \qty[group-digits=none]{0.00959}{\percent}, and an interquartile range of \qty[group-digits=none]{0.00539}{\percent}.
This places \qty{75}{\percent} of the historical forecasts below \qty[group-digits=none]{0.01262}{\percent} of error, with values as low as \qty[group-digits=none]{0.00191}{\percent}.
The overall performance of the theta model is good for the majority of the \glspl{ro}.
The hyperparameters used for the backtest evaluation have been tuned for a single \gls{ro}, for which the theta model yields very low errors.
While this set of parameters shows a good generalization across devices, further tuning per oscillator can assist in reducing the \gls{mape} values over the entire data set.

 \section{Conclusions and Outlook}\label{sec:conclusion}

In this work, we performed a long-term study of the degradation in propagation delay over a fleet of \num{298} naturally aged \glspl{fpga} that operate at the \gls{xfel} particle accelerator.
With the deployment of a measurement module, we collected field data from ring oscillators on the devices during maintenance and operation time periods.

By studying measurements from machine shutdowns six months apart, we quantified the effects of degradation mechanisms on the hardware during operation.
Our analyses revealed that the chips underwent a significant ageing over this period, which reduced the oscillation frequencies by a median of \qty{0.0496}{\percent} and a maximum in the order of \qty{0.25}{\percent}.
Aggregating frequency shifts per device showed a noticeable increase in propagation delay across all \glspl{fpga}.
We evaluated early detection of anomalous behaviour based on the estimation of abnormal degradation, by calculating a modified Z-score for each device based on \gls{ro} frequency shifts.

Our statistical analysis on data from continuous monitoring showed strong indicators of generalized degradation across devices, with a median reduction in oscillation frequency of \qty{0.064}{\percent} over \num{280}~days of study.
The calculated statistical metrics suggest a steady negative correlation between time passage and \gls{ro} frequency, which provides a better understanding of the behaviour of hardware ageing in the field.

The large scale of our study allowed us to spatially locate degradation levels of propagation delay over the \gls{fpga} fabric.
By correlating measurement data with knowledge of the placement of logic cores on the device during operation, we identified the impact of various circuits on the underlying hardware.
This type of analysis provides detailed and valuable insights that can be exploited by developers to implement graceful degradation and self-healing techniques.

By exploring \gls{ml} forecasting models and configurations, we achieved estimations of future frequency trends based on monitoring data with mean errors of \qty{0.002}{\percent} when observing up to a forecast horizon of \num{60}~days, evaluated over a period of \num{100}~days.
We further explored the aptitude of certain models for predictive maintenance by applying the backtest technique on historical data and analysed the trade-offs between training length, forecast horizon, and performance.

We envision that this work fosters further research about the degradation of embedded devices based on real-world data from naturally aged hardware and large-scale deployments.
In the future, we will integrate the developed monitoring and forecasting techniques into an automated predictive maintenance system for distributed facilities such as the \gls{xfel}.
Moreover, we will explore the application of similar monitoring principles to microcontrollers and their integration with an open source embedded operating system, such as RIOT~\cite{bghkl-rosos-18}.
 
\ifanonymous
\else
\section*{Acknowledgement}
We acknowledge the support by DASHH (Data Science in Hamburg - HELMHOLTZ Graduate School for the Structure of Matter) with the Grant-No. HIDSS-0002,
as well as the support of the Federal Ministry of Education and Research with Grant C-ray4edge.
 \fi

\bibliographystyle{IEEEtran}

\label{lastpage}

\end{document}